\documentclass[sigconf]{acmart}
\AtBeginDocument{%
  \providecommand\BibTeX{{%
    \normalfont B\kern-0.5em{\scshape i\kern-0.25em b}\kern-0.8em\TeX}}}

\copyrightyear{}
\acmYear{}
\acmDOI{}
\acmISBN{}

\acmConference[CIDR'22, January 9-12, 2022]{}{Chaminade CA}{USA.}

\makeatletter
    \def\@copyrightpermission{%
       This paper is published under the Creative Commons Attribution~4.0
       International (CC-BY~4.0) license. Authors reserve their rights to
       disseminate the work on their personal and corporate Web sites with
       the appropriate attribution, provided that you attribute the original work to the authors and CIDR 2022. 12th Annual Conference on Innovative Data Systems Research (CIDR ’22). January 9-12, 2022, Chaminade, USA.
}
\makeatother

\usepackage{xcolor}
\usepackage{mathptmx}
\usepackage{graphicx}
\usepackage{times}
\usepackage{listings}
\usepackage{courier}
\usepackage{amsmath}
\usepackage{xspace}
\usepackage{booktabs}
\usepackage{enumitem}
\usepackage{soul}

\usepackage{multicol}
\usepackage{multirow}
\usepackage{tabularx}
\usepackage{xcolor}
\usepackage[T1]{fontenc}

\usepackage{wasysym}
\usepackage{url}
\usepackage{hyperref}
\usepackage{titlesec}
\usepackage{enumitem}
\usepackage{subfigure}
\usepackage{tikz}

\usepackage{listings}
\usepackage{caption}
\usepackage{tabu}

\newif\ifnotes
\notestrue

\newcommand{\subheadi}[1]{\noindent\textit{#1}}
\newcommand{\ignore}[1]{}
\newcommand{\code}[1]{\texttt{#1}}
\newcommand{\system}{{\sc SystemD}\xspace}

\makeatletter
\renewcommand{\paragraph}{%
  \@startsection{paragraph}{4}%
  {\z@}{.4ex \@plus .4ex \@minus .2ex}{-.5em}%
  {\normalfont\normalsize\bfseries}%
}
\makeatother

\begin{document}

\title{Augmenting Decision Making via Interactive What-If Analysis}

\author{Sneha Gathani}
\affiliation{%
  \institution{Sigma Computing}
  \city{San Francisco}
  \country{USA}
  }
\email{sneha@sigmacomputing.com}
\additionalaffiliation{%
  \institution{University of Maryland}
  \city{College Park}
  \country{USA}
}

\author{Madelon Hulsebos}
\affiliation{%
  \institution{Sigma Computing}
  \city{San Francisco}
  \country{USA}
}
\email{madelon@sigmacomputing.com}
\additionalaffiliation{%
  \institution{University of Amsterdam}
  \city{Amsterdam}
  \country{Netherlands}
}

\author{James Gale}
\affiliation{%
  \institution{Sigma Computing}
  \city{San Francisco}
  \country{USA}
}
\email{jlg@sigmacomputing.com}

\author{Peter J. Haas}
\affiliation{%
  \institution{University of Massachusetts}
  \city{Amherst}
  \country{USA}
}
\email{phaas@cs.umass.edu}

\author{\c{C}a\u{g}atay Demiralp}
\affiliation{%
  \institution{Sigma Computing}
  \city{San Francisco}
  \country{USA}
}
\email{cagatay@sigmacomputing.com}

%

\begin{abstract}
The fundamental goal of business data analysis is to improve business decisions using data. Business users often make decisions to achieve key performance indicators (KPIs) such as increasing customer retention or sales, or decreasing costs. To discover the relationship between data attributes hypothesized to be drivers and those corresponding to KPIs of interest, business users currently need to perform lengthy exploratory analyses. This involves considering multitudes of combinations and scenarios and performing slicing, dicing, and transformations on the data accordingly, e.g., analyzing customer retention across quarters of the year or suggesting optimal media channels across strata of customers. However, the increasing complexity of datasets combined with the cognitive limitations of humans makes it challenging to carry over multiple hypotheses, even for simple datasets. Therefore mentally performing such analyses is hard. Existing commercial tools either provide partial solutions whose effectiveness remains unclear or fail to cater to business users altogether.

Here we argue for four functionalities that we believe are necessary to enable business users to interactively learn and reason about the relationships (functions) between sets of data attributes thereby facilitating data-driven decision making. We implement these functionalities in \system, an interactive visual data analysis system enabling business users to experiment with the data by asking \emph{what-if} questions. We evaluate the system through three business use cases: marketing mix modeling, customer retention analysis, and deal closing analysis, and report on feedback from multiple business users. Overall, business users find the \system functionalities highly useful for quick testing and validation of their hypotheses around their KPIs of interest, clearly addressing their unmet analysis needs. Their feedback also suggests that the user experience design can be enhanced to further improve the understandability of these functionalities. 
\end{abstract} 

\maketitle

\section{Introduction}\label{sec:introduction}
Interactive visual data analysis systems targeting business users (e.g., sales, marketing, product, or operations managers) aim to help users make data-driven decisions. A basic yet overarching question related to this goal is what do business users essentially try to achieve with data analysis? Specifically, what should an enterprise data analysis system designed for business users be optimizing for? Answering this question is critical for operationalizing business data analysis. 

The current view on interactive visual data analysis has been primarily shaped by John Tukey's emphasis on exploratory data analysis (EDA)~\cite{tukey1962future}. Tukey, a prodigious figure with wide-ranging contributions~\cite{brillinger2002johnwtukey} to statistics and beyond, considered data analysis in two stages: exploratory analysis, which he likened to the detective work of collecting evidence in an investigation, and confirmatory analysis (e.g., statistical hypothesis testing), analogous to the trial step of the investigation, where the validity and the strength of collected evidence need to be proved for a judge or jury. Tukey highlighted the importance of exploratory analysis and the use of graphics (visualizations) to that end, which had been overlooked by the statistical community of his time. The developments in the last two decades, including the wider adoption of interactive visualizations in data analysis and the success of commercial as well as open-source EDA tools, demonstrated the value of Tukey's perspective on data analysis. Paradoxically, this success also caused a tunnel vision that frequently turned exploratory analysis into the end itself by overly focusing on visual pattern exploration and qualitative hypothesis generation~\cite{hullman2021designing}. This skewed focus limits existing systems in helping business users to effectively complete their data analysis goals.  

Consider these user questions: \textit{how can I best use my \$200K marketing budget across advertisement channels? Which activities should I prioritize to increase my deal closing rate? How much do free trials help with acquiring customers? What drives the increase in revenue?} These decision-leading questions are, at best, difficult and time-consuming to answer through existing interactive exploratory analysis systems or data science tools by business users, who typically have no time for or background in coding, statistical analysis, or algorithmic modeling. These questions also hint at what we consider to be the purpose of business data analysis.

\begin{figure*}
    \centering
\includegraphics[width=1\linewidth]{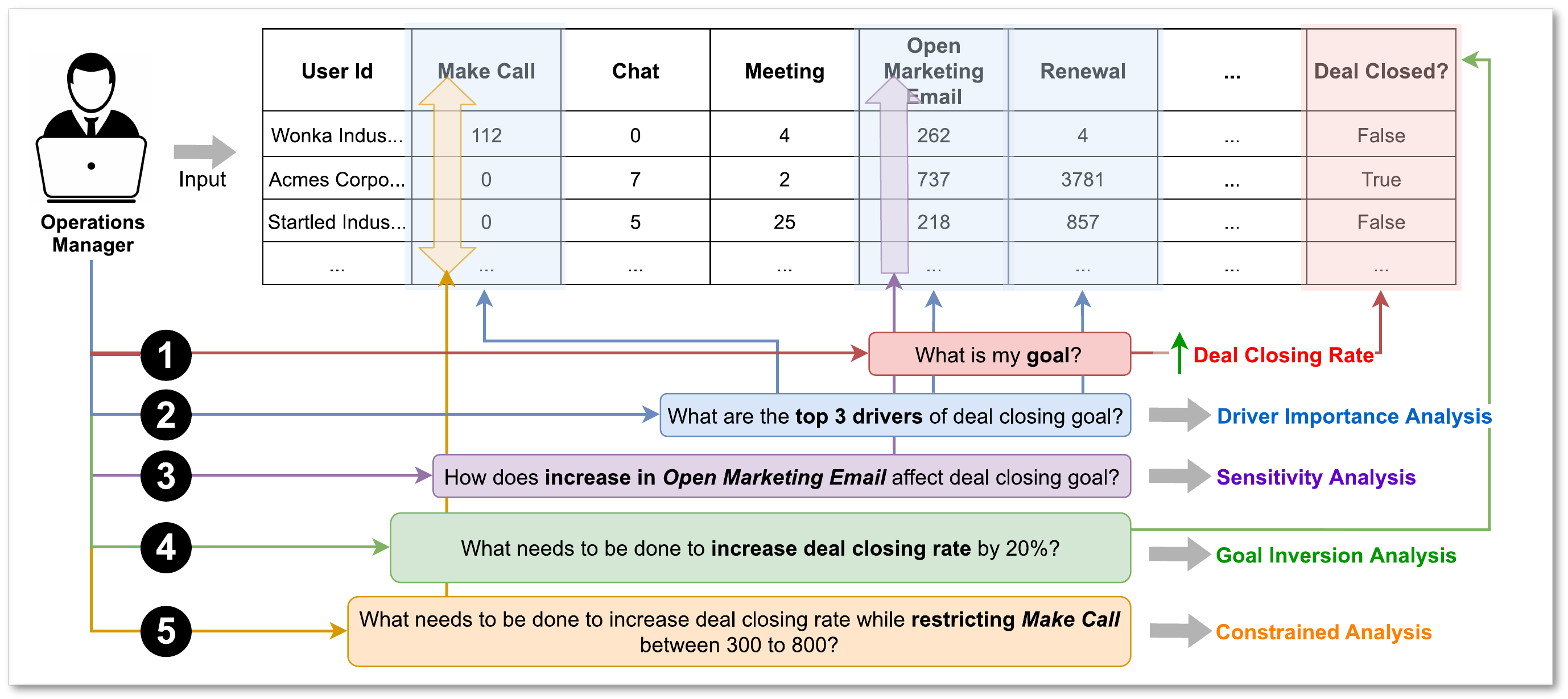}
    \caption{Four functionalities proposed by \system to augment business users beyond exploratory analysis in decision making: Driver Importance Analysis, Sensitivity Analysis, Goal Inversion (Seeking) Analysis, and Constrained Analysis.}
    \label{fig:teaser}
\end{figure*}

\paragraph{Purpose of Business Data Analysis.} The fundamental goal of business data analysis is to improve business decisions. 
This is done by understanding the relationship between two sets of variables--input or independent variables which are hypothesized to be potential drivers and output or dependent variables (often a single variable) that are business key performance indicators (KPIs) and are hypothesized to be dependent on the input variables. For the remainder of the paper, we will refer to the input or independent variables as \textit{drivers} and the output or dependent variables as \textit{KPI}s.

\paragraph{Challenges.} While interactive exploratory data analysis is useful, it is not sufficient for effectively carrying out the fundamental task above. There are four basic challenges.

\begin{itemize}[itemsep=0.5ex,leftmargin=0em, topsep=0.5ex]
\item[]\subheadi{\hspace{3mm}Human Cognition.} 
Limitations of human working memory and cognitive overload due to time pressures and data complexity limit the user's ability to effectively run what-if scenarios, without getting help for rigorously following effective methods to generate, manage, and evaluate hypotheses. Confirmation bias~\cite{nickerson1998confirmation}, our tendency to fit the evidence to existing expectations, makes it hard to explore insights in an unbiased and rigorous manner~\cite{pirolli2005sensemaking,tversky1974judgment}. Thus, people often fail to focus on the most relevant evidence while sufficiently attending to hypotheses' disconfirmation~\cite{nickerson1998confirmation}.
\item[]\subheadi{\hspace{3mm}Interactive Exploratory Analysis.} Interactive direct manipulation as a querying and data transformation paradigm does not scale well for learning relations (functions) between drivers and KPIs, often requiring a large number of transformations as well as slicing and dicing combinations.
\item[]\subheadi{\hspace{3mm}Data Scale and Complexity.}
Increased data sizes and complexities exacerbate the two challenges above, easily turning the fundamental task of business data analysis into a daunting process if not impossible. Note that data in enterprise databases are constantly updated and appended, which cloud computing has been making easier and cheaper. Yesterday’s feasible choices or decisions informed by data can easily be suboptimal or infeasible today due to updates or the availability of new data. So, large complex dynamic data put additional pressure on human cognition and working memory.
\item[]\subheadi{\hspace{3mm}Dead Data.} 
Even if domain expert business users can build mental models between potential drivers and KPIs for a snapshot of data, there is no easy way for them to probe into, reason about, and run scenarios over to stress-test and utilize this mental model for decision making~\cite{haas2011data}. Mental models built out of exploratory analyses don’t lend themselves to simulation or scenario modeling based on hypothetical data, which is necessary for what-if analysis.
\end{itemize}

\paragraph{Elements of Business Data Analysis.} 
What are the elements of the fundamental purpose of business data analysis introduced above? What does improving decisions mean? Based on our conversations with business users, we distill the following observations.
\begin{itemize}[itemsep=0.5ex,leftmargin=0em, topsep=0.5ex]
\item[]\subheadi{\hspace{3mm}Improve Decisions.} The goal of data analysis is to improve decisions based on data. An improved decision---an effective operationalization of insights---manifests itself differently in different domains and use cases. It can be increased sales, reduced cost, increased customer retention rate, reduced churn rate, reduced customer acquisition cost, and so on. Business users mean, well, business.
\item[]\subheadi{\hspace{3mm}Understand Driver-KPI Relationship.} Improving decisions requires users to learn, manually (mentally) or otherwise, the relationships  (functions) between drivers in their data and KPIs on which their business objectives are based.
\item[]\subheadi{\hspace{3mm}Use Data and Domain Expertise.} Decision-making is an interplay between data and domain knowledge, including common sense. Neither data nor domain knowledge (expertise), which business users possess, is sufficient by itself for improved outcomes. 
\item[]\subheadi{\hspace{3mm}Value of BI Systems.} The value of an enterprise data analysis or business intelligence (BI) system is its added value in effectively enabling improved decision-making using data and domain knowledge.
\end{itemize}

\paragraph{Desiderata for Systems.} Based on the observations above and earlier work~\cite{cavallo2018visual}, we propose that enterprise data analysis systems integrate four interactive functionalities (\autoref{fig:teaser}) to augment business users in decision making, going beyond exploratory analysis.
\begin{itemize}[itemsep=0.5ex,leftmargin=0em, topsep=0.5ex]
\item[]\subheadi{\hspace{3mm}Driver Importance Analysis.} Enables users to implicitly learn functions (models) allowing them to understand the relationships between drivers (input) and KPIs (output), along with the artifacts of these learned relationships such as the relative importance of various drivers and their interactions in predicting the KPI outcomes.
\item[]\subheadi{\hspace{3mm}Sensitivity Analysis.} Enables users to dynamically evaluate learned relationships for arbitrary driver values and observe the changes in KPI values. This helps users build their intuition about how their business works in a hands-on manner. 
  To this end, systems should help users to experiment with the drivers by interactively perturbing (increasing or decreasing) their values and observing the effects on the KPI values.
\item[]\subheadi{\hspace{3mm}Goal Inversion (Seeking) Analysis.} Enables users to interactively set goals such as specific target values or optimization objectives (maximization or minimization) for the KPIs and observe multiple scenarios on how the driver values need to change to achieve the desired goals. For example, systems should provide recommendations for changes needed in driver values to achieve user-specified KPI goals such as doubling the revenue or minimizing the churn rate.
\item[]\subheadi{\hspace{3mm}Constrained Analysis.} Allows users to interactively set constraints or conditions over how the learned functions (models) are evaluated or inverted, enabling users to incorporate their domain knowledge such as business constraints and common sense to regulate these functions. This also enables users to quickly generate and evaluate multiple scenarios under various conditions specified by constraints. For instance, systems should enable users to set constraints (e.g., boundary, equality, or inequality) on one or more drivers and run the goal-seeking analysis to provide 
driver values satisfying user constraints. Note that there is rarely a single best context-free decision for improving a KPI goal; instead, there are often multiple feasible choices with dynamic costs and trade-offs bound to decision paths. Systems should enable rapid discovery as well as management and tracking of these choices (options), making them first-class citizens of data analysis. 
\end{itemize} 

We operationalize the desiderata above through an interactive visual data analysis system called \system~\footnote{\url{https://tinyurl.com/SigmaSystemD}}. The system enables users to experiment with their data and understand relationships between drivers (input) and KPIs of interest (output) through what-if scenarios. We evaluate \system using three common business use cases: (1) marketing mix modeling, (2) customer retention analysis, and (3) deal closing analysis with real-world business users such as sales, marketing, operations, accounts, and campaign managers.

In summary we make the following contributions:
\begin{itemize}[itemsep=0ex, leftmargin=1.85em, topsep=0ex]
    \item We revisit the fundamental goal of business analysis to be making data-driven decisions via the ability to understand the relationship between drivers (input) and KPIs (output).
    \item We argue that the four functionalities we propose,  \textit{driver importance analysis, sensitivity analysis, goal inversion (seeking) analysis,} and \textit{constrained analysis}, are necessary for effectively helping business users to make data-driven decisions.
    \item We implement \system, an interactive visual data analysis system incorporating these functionalities, providing a coherent workflow and end-to-end solution for business users to improve their decisions.
    \item We evaluate \system through three business use cases to test the applicability and perceived value of the system.
\end{itemize}

\section{SystemD}
\label{sec:interface_design}

\begin{figure*}[t]
    \centering
    \includegraphics[width=\textwidth]{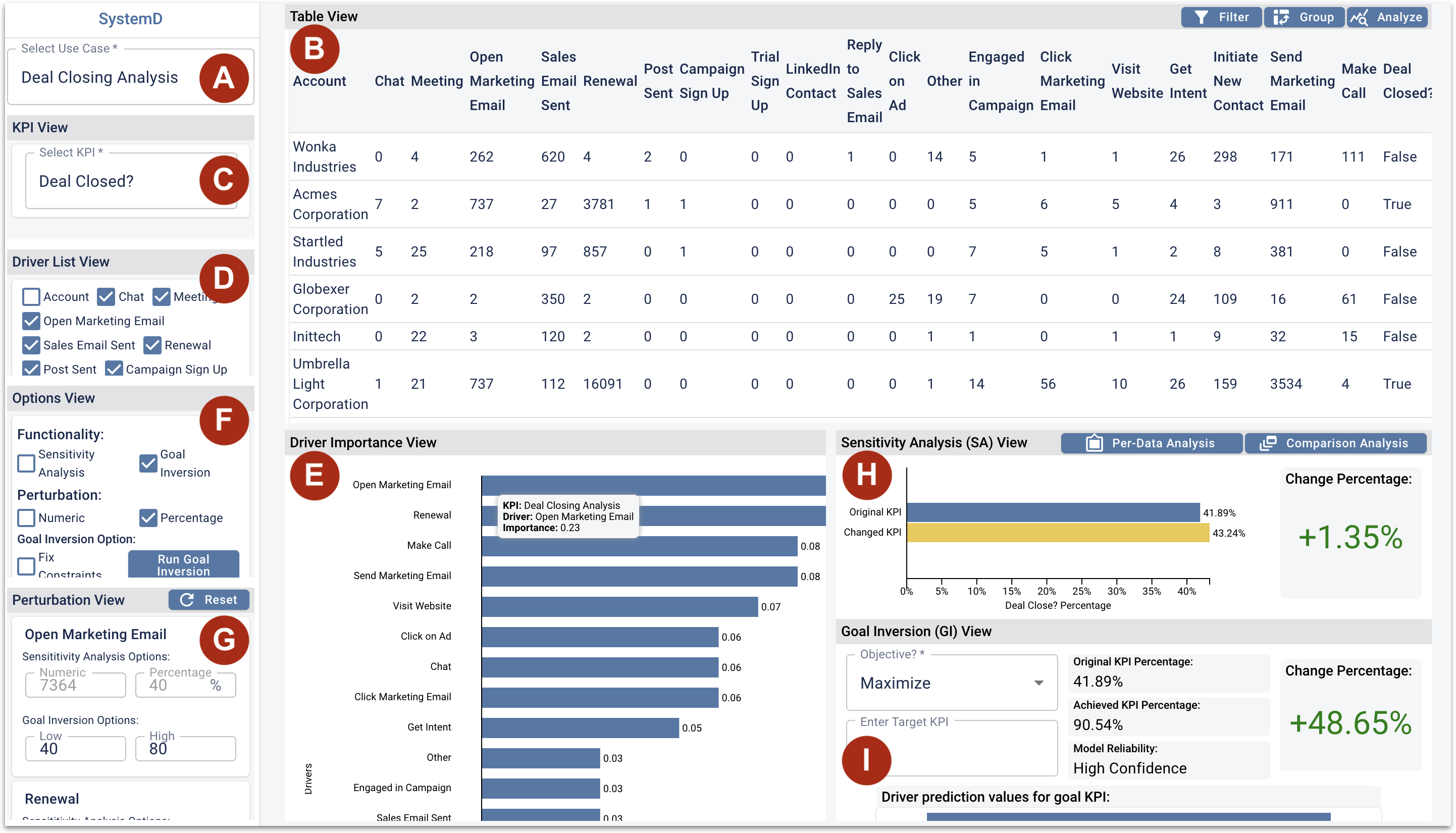}
    \caption{\system user interface showing a snapshot from the deal closing analysis use case. It operationalizes four functionalities we deem necessary for augmenting decision making in data analysis systems. See Section~\ref{sec:interface_design} for details on each of the annotated views.}
    \label{fig:system}
\end{figure*}

\system has a client-server architecture. The frontend client is responsible for rendering analysis views, managing user interactions, communicating with the backend server to fetch requested data, and (re-)rendering views accordingly. The backend server runs machine learning models to predict KPI objective values and packs them into efficient JSON data structures to send to the client in response to user interactions. We train two widely used models: linear regression models when the KPI objective is a continuous variable (e.g., sales) and classifiers when the KPI objective is a discrete variable (e.g., customer retained after 6 months or not, deal closed or not) to make predictions. Now, we briefly describe the views and functionalities of \system as annotated in \autoref{fig:system}.

\paragraph{(A) Use Case Selection.} Users can select from three common business use cases supported currently. \autoref{fig:system} shows a snapshot of \system during the deal closing analysis use case.

\paragraph{(B) Table View.} On selection of the use case, the corresponding dataset is tabulated in this view. The columns in this view show the drivers and KPI of the dataset. \ignore{Currently, we expect the data pre-processing and cleaning to be performed before it is loaded to the system for analysis.}

In the deal closing analysis use case dataset, every row represents a prospective customer and every column represents the counts of activities corresponding to the prospect such as \textit{Chat}s, \textit{Meeting}s attended, etc. Each prospect is also labeled as converted to a customer or not.

\paragraph{(C) KPI Selection View.} Users can select one of the variables in the dataset to be the KPI objective using this view.

For the use case, users are interested in whether prospects close as customers, hence choose the KPI to be \textit{Deal Closed?}.
  
\paragraph{(D) Driver List View.} This view allows users to filter out drivers they aren't interested in analyzing.

Users want to understand the relationship between all activities of prospects and the average deal closing rate in the use case. Hence, they keep all variables selected and deselect only the \textit{Account} textual variables to enable model training and prediction.

\paragraph{(E) Driver Importance View.} This view characterizes the first functionality of \system, the \textit{driver importance analysis}. Users are interested in recognizing the drivers that drive the KPI. This view illustrates an interactive horizontal bar chart informing the users of the drivers (Y axis) of the KPI in decreasing order of importance (X axis) thereby guiding them to take actions in the appropriate directions.

We use Scikit-learn~\cite{pedregosa2011scikit} to train machine learning models that predict KPI values: linear regression models for continuous KPIs and random forest classifiers for discrete KPIs. We choose the driver importance values to be the linear-regression coefficients and the random-forest feature importances because they are relatively easier for users to understand. However, we verify the importances using traditional measures such as Shapley, Pearson, and Spearman rank~\cite{roth1988shapley, benesty2009pearson, zar2005spearman} to ensure that the model coefficients are not misleading. The importance values range between -1 and 1 with extremes showing high negative and positive importance to the KPI respectively while closer to 0 shows low importance to the KPI.

Driver importance analysis for our use case shows that the three most important drivers of the deal-closing KPI are \textit{Open Marketing Email}, \textit{Renewal} and \textit{Call} while the three least important drivers are \textit{LinkedIn Contact}, \textit{Initiate New Contact} and \textit{Meeting}.

\paragraph{(F) Options View.} This view provides options for types of analyses to perform and types of perturbations to apply to the original dataset. Users can check corresponding boxes to perform either sensitivity analysis or goal inversion analysis. The system supports two perturbation options: absolute and percentage values. The deal closing analysis use case illustrates outputs of both analyses using percentage perturbations.

\paragraph{(G) Perturbation View.} Users can specify how much to perturb the original dataset using this view. For each of the drivers, they can set absolute or percentage perturbation magnitudes to perform sensitivity analysis and low and high constraints for goal inversion analysis.

Perturbations are made to each of the data points in the dataset; for example, a 40\% increase on \textit{Open Marketing Email} means increasing the marketing emails opened for every prospect by 40\%.

\paragraph{(H) Sensitivity Analysis View.} \system helps users understand the relationship between drivers and the selected KPI using its second functionality of \textit{sensitivity analysis}. Users can experiment with the drivers by perturbing the original data --- see (G) above --- and observing the effects on the KPI.
This view displays two bars: a static blue bar shows the KPI achieved on the original dataset and an interactive yellow bar shows the KPI achieved on the perturbed dataset. The change between KPI achieved on user perturbed data and the original data is explicitly shown to indicate the up-lift (positive, shown in green) and down-lift (negative, shown in red). Every perturbation re-runs the model prediction to re-calculate the KPI value.

To ease driver experimentation, \system provides two additional sensitivity analysis features. To view sensitivity analysis in its entirety and compare KPI trends over all drivers, the \textit{comparison analysis} feature shows the KPI achieved for every driver individually across a range of perturbations specified by the user. To allow users to drill down and experiment with individual data points, the \textit{per-data analysis} feature lets users select a data point, perturb data of the selected data point and view effects on its KPI value.

The use case shows a 40\% perturbation made to the \textit{Open Marketing Email} predictor, which results in the deal-closing rate rising to 43.24\% (yellow bar), an up-lift of 1.35\% from that on original data.

\paragraph{(I) Goal Inversion View.} The \textit{goal inversion} functionality helps users to plan future actions that will help them achieve their KPI goals. Users can choose to freely optimize (minimize or maximize) the KPI or specify a desired KPI value they wish to achieve and run the model to provide values of the drivers that will achieve the goal KPI.

\system uses Scikit-Optimize's Bayesian optimizer~\cite{louppe2017bayesian} to learn values of the drivers that attain the desired KPI value (maximum, minimum, or target). The view returns the best KPI attainable, the confidence of the model used, and a set (not necessarily unique) of driver values that achieve the user-specified KPI goal.

In practice, it is not always feasible for users to take the actions recommended by freely optimized goal inversion, such as when recommendations are not in line with domain knowledge or when optimal driver values exceed a budget, and so on. To obtain realistic recommendations, \system provides the final functionality of \textit{constrained analysis}. This functionality allows users to specify low and high constraints on one or more drivers and have the model recommend optimal KPI and driver values that are within the specified range.

The use case shows a \textit{constrained analysis} in which users constrain the perturbation of the \textit{Open Marketing Email} driver to be an increase between 40\% to 80\% as shown in (G). The resulting maximal deal-closing-rate KPI equals 90.54\%, which is an up-lift of 48.65\% from the KPI achieved on the original data.
\section{Evaluation}\label{sec:evaluation}

\begin{table*}[t]
\caption{Pre-study, system usability (elicited on a Likert scale of 5), and open-ended questions used for evaluating \system.}
\vspace{-3mm}
\resizebox{\textwidth}{!}{
\begin{tabular}{cl}
\toprule
\textbf{Category} & \multicolumn{1}{c}{\textbf{Questions}} \\ \midrule
& \textit{Can you describe the kind of data you use?} \\
 & \textit{What is the intent of using the data?} \\
 & \textit{Given the data, what would you be most interested in analyzing?} \\
 & \textit{What is the purpose behind interest in the analysis of the data?} \\
 & \textit{\begin{tabular}[c]{@{}l@{}}Consider you are interested in sales (U1)/retention rate (U2)/deal closing rate (U3), can you describe what analysis would you perform to make\\ decisions on investing in the right channels (U1)/increasing the retention rate (U2)/increasing deal closing rate (U3)?\end{tabular}} \\
 & \textit{Which tools do you use typically to perform the analyses you described?} \\
 & \textit{How easy or hard would you say it is for you to analyze the data and make a decision?} \\
 & \textit{How much time would you approximately take to come up with a hypothesis and make a decision based on that?} \\
\multirow{-9}{*}{Pre-study} & \textit{What strategies do you use to evaluate whether analyses results match your expected hypotheses (via your domain knowledge and/or experience)?} \\
\addlinespace
\hline
\addlinespace
 & \textit{The functionalities of SystemD are useful in understanding the behavior of the data better.} \\
& \textit{The functionalities of SystemD are useful in making optimal decisions.} \\
 & \textit{The interactions with SystemD are intuitive.} \\
 & \textit{Most users would learn to use SystemD very quickly.} \\
 & \textit{Various functionalities of SystemD are well-integrated.}
 \\
 & \textit{\begin{tabular}[c]{@{}l@{}}Compared to your process of analysis and current tools you use on a daily basis for making decisions (as described initially), how useful do you see\\ SystemD helping you for the same tasks?\end{tabular}} \\
\multirow{-8}{*}{\begin{tabular}[c]{@{}c@{}}System usability\\
\end{tabular}} & \textit{Use SystemD in my daily work.} \\
\addlinespace
\hline
\addlinespace
 & \textit{\begin{tabular}[c]{@{}l@{}}Compared to your process of analysis and current tools you use on a daily basis for making decisions (as described initially), how useful do you see\\ SystemD helping you for the same tasks? Explain why.\end{tabular}} \\
 & \textit{How useful is SystemD for making decisions that optimize interesting metrics (KPIs) in comparison to current tools? Explain why.} \\
 & \textit{\begin{tabular}[c]{@{}l@{}}List the most useful functionalities or features from most useful to least useful (Driver Importance Analysis, Sensitivity Analysis, Goal Inversion\\ (Seeking) Analysis, Constrained Analysis).\end{tabular}} \\
 & \textit{Which additional functionalities or features would become a more effective system to make decisions in SystemD?} \\
\multirow{-5}{*}{Open-ended} & \textit{What would be your concerns with the SystemD?} \\
\bottomrule
\end{tabular}
}
\label{tab:evaluation_questions}
\end{table*}

We evaluate \system through three business use cases to gather feedback on its interactive functionalities. 

\paragraph{Participants and Protocol.} We recruited five business users who play various roles at Sigma Computing and have a minimum experience level of four years. Three business users---a marketing manager, a campaign manager, and an account manager---participated in the marketing mix modeling use case. A product manager and a sales manager participated in the customer retention analysis and the deal closing analysis use case respectively. Although we recruited Sigma employees as our business users in order to facilitate our evaluation process, these employees are nonetheless fair representatives of the types of business users who daily perform analyses and make decisions.

Participants were briefed on the interview goals and structure. They were then asked to share analysis intent, workflows, current tools and challenges of their daily analysis in a pre-interview questionnaire, both to set the stage for the experiment and to let us understand in detail their context for decision-making using data.
Then we shared our screen and demoed the system to the participants to explain the four functionalities. The three kinds of business users were each shown the system using their respective use case of interest so as to match their daily workflows. We demoed the system using certain interesting examples which we previously found while experimenting with the dataset, but we highly encouraged the participants to also guide us on the experiments they wanted to perform. For example, the product manager in the retention rate analysis use case explicitly asked us to remove an obvious predictor and perform the functionalities again.
After the interview, participants were asked to provide feedback by completing a system usability questionnaire. Participants rated their agreement with system usability questions
using a Likert scale of 1 (strongly disagree) to 5 (strongly agree). Additionally, we asked participants to provide us with more open-ended feedback on the features of, and concerns about, \system. The set of questions from the pre-interview questionnaire, system usability questions and open-ended feedback questions are listed in \autoref{tab:evaluation_questions}. Interviews typically lasted 45 minutes. We used a think-aloud protocol to encourage participants to share their thought process simultaneously as they got results from \system. We recorded audio and captured screen for later review.

\paragraph{Use Cases.} Here we briefly describe the three business use cases.
\begin{itemize}[itemsep=0.5ex,leftmargin=0em, topsep=0.5ex]
\item[]\subheadi{\hspace{3mm}U1: Marketing Mix Modeling.} Marketing Mix Modeling is a technique which helps quantify the impact of several marketing inputs on sales, market share, etc. With the rapid increase of new media, there has been an increased need for marketeers to optimize their strategies that maximize KPIs such as return of investment, sales, profit, etc. across multiple channels.

To allow Sigma's marketing, campaign and account managers to use \system for making decisions that optimize investments per media channel, we used a dataset describing investments made over a period of 6 months on 5 media channels (Internet, Facebook, YouTube, TV and Radio) and corresponding sales achieved per day. The participants guided the analysis for understanding the behavior between media channels and sales KPI and made decisions on which channel investments should increase or decrease to maximize sales.

\item[]\subheadi{\hspace{3mm}U2: Customer Retention Analysis.} Operations and product managers are interested in retaining existing customers because it is much easier, both in terms of effort and cost, to retain old customers than get new ones. To build strategies that retain customers, business users analyze large datasets of customer interactions with the product and their own activities with the customers that are collected by CRM and support systems. Examples of such activities are using help chat, opening new document, adding a visualization, etc. Additionally, business users often form hypothesis of additional features, or \textit{formulas} that may drive customer retention such as customers attending 2+ demo meetings in the first two weeks, and so forth.

Sigma's product manager was interested in finding customer activities that would maximize retention after six months. Sigma's multi-touch attribution dataset was used for analysis; it consists of a customer's activities and product manager's hypothesis formulas such as pivoting on data, performing join operation, using 3+ formulas in two weeks, etc., during the last six months, along with a label indicating whether the customer was retained after six months. The product manager used \system to optimize the six-month-retained KPI and understand the activities that lead to increased six month retention.

\item[]\subheadi{\hspace{3mm}U3: Deal Closing Analysis.} The approach to selling a product begins with initiating contact with a prospective customer and ends after a salesperson closes the deal. This process consists of multiple stages of back and forth communication between the prospects and the sales teams. Similar to the customer retention analysis use case, sales managers use a prospect's activities with the product, as well as the sales and marketing teams communication activities, to predict whether prospects will close to customers. Examples of such activities are prospects signing up for trials or demos, participating in campaigns, attending scheduled meetings, and so on.

The sales manager at Sigma wanted to analyze prospect data from the past few months similar to the data described in Section~\ref{sec:interface_design} to answer questions such as ``what activities lead to the conversion of more prospects?'' and ``what is the ideal customer journey formula for Sigma?''.
\end{itemize}
\section{Results}
\label{sec:results}
\begin{figure}[t]
\centering
\includegraphics[width=\columnwidth]{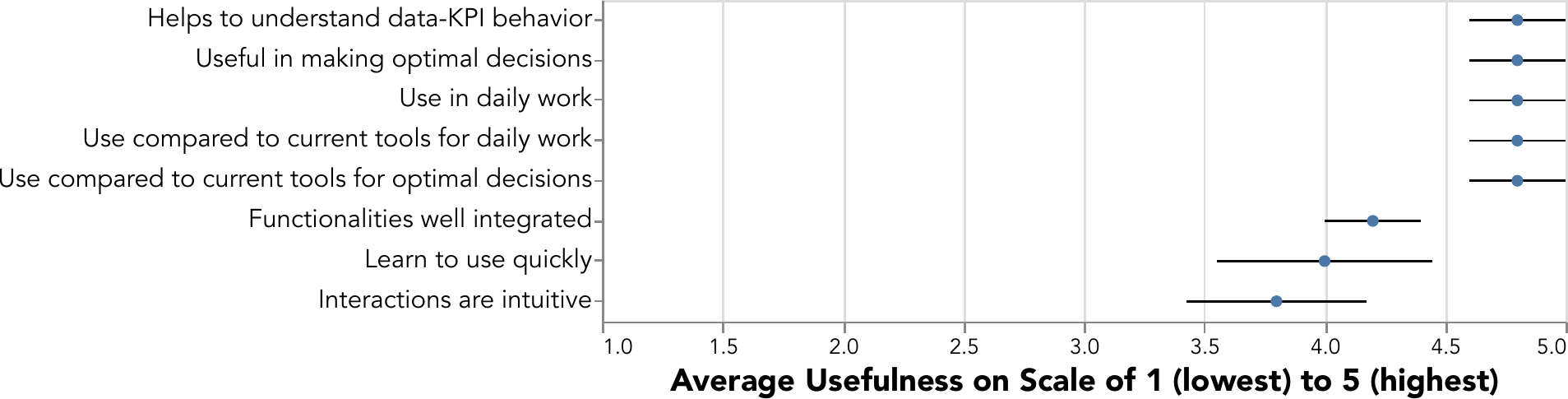}
\caption{Feedback elicited through responses to usability questions on a Likert scale of 5. Participants found \system to be highly useful for understanding data-KPI behavior, making optimal decisions, and adopting it in their daily work. On the other hand, they rated its interactions to be intuitive to a lesser degree.} 
\label{fig:evaluation}
\end{figure}

We now discuss the qualitative feedback we collected from Sigma business users on \system (\autoref{fig:evaluation}). 

\paragraph{Understanding Driver-KPI Relationship.} We found that participants were able to quickly understand the dynamics between drivers and their KPI objective and make optimal decisions using \system. When describing her prior experiences with campaign analysis, the campaign manager said \textit{``[right now] we have to manually set accurate targets for our campaigns looking at the historical data and intuitively figure out the media channels [that] have the best [chance at] converting leads [of users] into actual customers''} and noted that \system is \textit{``definitely much more actionable!''}. We found that all other participants shared similar sentiments. For example, while describing the challenges with current tools used such as reporting tools and spreadsheets, the marketing manager remarked that her \textit{``team consists of only marketers and not technical engineers or data scientists, making them really struggle ... to really see what is going on in spreadsheets and this [\system would be] ... very exciting to perform ... [such] analyses''}.

\paragraph{Value For Business Users.} All five participants saw very strong value in using \system as part of their daily workflows. Participants shared that they mostly used reporting tools such as Sigma, Microsoft Excel or Salesforce for their analysis. They also shared that they did not think any of these tools thrived in statistical analysis, making them unaware of any machine learning or predictive analytics functionalities in the tools. They repeatedly found the interactive statistical analysis of \system beneficial as it could save them from the intuitive and time-consuming trial and error strategy of hypothesizing next steps, performing them and waiting for three to six months to see the results. The product manager liked the ability of the interactive sensitivity analysis functionality to help her narrow this expensive strategy which \textit{``... is not something that [she] ... is easily able to do right now''}.

Business users in two use cases were interested in working directly with the \system team to fine-tune the system for their individual needs immediately. For example, after walking through the analysis demo, the accounts manager exclaimed that \textit{``[she] wanted to get access to [\system\/] now!!!''} and asked \textit{``whether we [her team] could access it ourselves [themselves]?''}. This showed the participant's interest in using \system right away.

\paragraph{Usefulness and Completeness.} When asked to rank the functionalities in terms of usefulness, we found that 3 of 5 participants ranked the driver importance analysis as being the most useful for their analysis while the other two participants ranked both sensitivity analysis and constraint-based analysis to be the most useful. All participants also ranked multiple functionalities second in usefulness and none being least useful. Thus, participants seemed to find all functionalities provided by \system useful. Most participants also found the functionalities well integrated and connected to one another.

\paragraph{Suggestions for Improvements.} 
Participants showed great interest in integrating \system with Sigma Worksheet\cite{gale2021sigma}, a visual query-builder interface, so they could add calculations for additional features or drivers to the dataset. This could also serve any need to slice, dice and drill to obtain the required analysis data, such as per customer-cohort or
prospect-stage analysis. Also, most participants needed clarification to understand the outputs of the functionalities, suggesting a need for better user experience workflows. We aim to resolve these concerns as we are starting to deploy \system into Sigma.

\section{Future Research Directions}
\label{sec:future_work}
Our study, along with our experience so far, confirms that helping business users make data-driven decisions implicitly using models is a difficult multifaceted problem with rich research and engineering challenges. We now discuss some of these challenges to indicate future research directions.

\paragraph{Communication and User Control.} Decision making involves two constituents: domain expert user(s) and models implicitly created through user interactions, as shown in \system. One challenge is how to best calculate and communicate the underlying model assumptions and confidences to users who have no background in statistics or data science? Another challenge is that some functionalities, such as goal inversion, may come across as ``magic,'' leading the users to form unrealistic expectations from the system. When the system fails to meet these expectations users may lose trust and become dissatisfied not just with the 
intelligent functionalities but also with the whole system. User trust is thus critical for automated methods, intelligent or otherwise~\cite{chuang2012interpretation}. Automated methods need to \emph{just work} and when they fail, they should fail consistently, while giving users a sense of control. This leads to the future problems of both modeling as well as system and user experience design. For instance, how can a user give feedback if she thinks that what the system recommended is incorrect? And, how do we retrieve this feedback and update our models and systems?
One solution is to use linear solvers~\cite{shang2021davos} which can scale but face challenges in providing fast real-time response when the data is large or when robust models are used for prediction. There is ample scope for future research.

\paragraph{Robustness.} The optimal solution from a given data-based model may be brittle: under small changes to the model or data, the solution may suddenly perform very poorly. This robustness problem is exacerbated by the fact that, for a given set of data, it is often the case that multiple models can reasonably explain the relationship between data and outcomes or objectives~\cite{breiman2001random, breiman2001statistical}. These different models may yield different rankings of driver importance as well as different solutions to optimization and goal-seeking problems. Therefore, data analysis systems designed to augment decision-making need to take the robustness of solutions and multiplicity of explanatory models into account. 

\paragraph{Interpretability vs Accuracy.} It is well-known that some models are simpler and easier to interpret while others are more accurate but difficult to explain. It is essential that we study which models to pick for our business users. Do we allow our users to have a say in this choice? 

\paragraph{Specification and Reuse.} In order to reuse the decision-making analyses process, it is important to have an editable specification of the experiments that \system supports. It is crucial to identify the right grammar for specifying these data experiments and enabling their interoperability with, e.g., other data science languages or platforms such as computational notebooks. An interesting direction is to integrate these experiments with SQL as many data analysis systems~\cite{cai2013simulation, brucato2020stochastic}, including Sigma, compile the data analysis intent of users into SQL queries. Therefore, development of a declarative specification language for \system is a potential future direction.

\paragraph{Conducting a Complete User Evaluation}
To rigorously learn the usefulness of SystemD beyond our initial evaluation, there is wide scope for conducting larger quantitative-based user studies. We want to understand various business user's existing decision workflows compared to when they use SystemD, and to compare quantitative factors such as whether a decision can be made, the amount of time needed to make the decision, and so forth. Another potential direction is comparing the functionalities of SystemD to the functionalities provided by other existing commercial tools such as Salesforce’s Einstein Discovery~\cite{einsteindiscovery} and Microsoft’s PowerBI~\cite{ferrari2016introducing}.
\section{Related Work}
\label{sec:related_work}
\paragraph{Model-based Analysis.} The interactive visual analytics~\cite{keim2008visual} literature has a plethora of prototype systems that incorporate models for data analysis. Typically these systems use unsupervised offline models to process data for feature extraction that subsequently augments interactive analysis. Less frequently, systems use online models to drive interactive analysis, for instance, the Praxis~\cite{cavallo2018visual} forward projection and backward projection for dimensionality-reduction-based data analysis. \system extends these two interactions to sensitivity analysis and goal inversion (seeking) analysis, respectively, for generalized model-driven relational data analysis for business users. 

\paragraph{Model Understanding.} Prior work introduces interactive systems for understanding models. 
Some systems have interactions similar to sensitivity analysis (e.g.,~\cite{wexler2019if, amershi2015modeltracker, zhao2018iforest}). Others use counterfactual explanations (e.g., \cite{hohman2019gamut,cheng2020dece,gomez2020vice,galhotra2021explaining}) on how to gain the desired prediction with minimal changes to the input, which is akin to our goal inversion (seeking) analysis. Unlike \system, the purpose of these approaches is to understand and debug machine learning models, not to support model-driven analysis. 

\paragraph{Data Farming.} In the context of simulation, the data farming literature~\cite{Sanchez20} emphasizes the value of a holistic understanding of a model (e.g., robustness properties, sensitivity) over simple optimization, which can yield high-reward but very brittle decisions. Unlike \system's hands-on approach, the emphasis in~\cite{Sanchez20} is on non-interactive visualization, and goal-inversion (seeking) is not explicitly supported.

\paragraph{Spreadsheets.} Spreadsheet applications generally enable optimization-based data analysis through add-ins and native functions. For example, Excel's \code{SOLVER}~\cite{exceldocsolver} and \code{GOAL SEEK}~\cite{exceldocgs} features allow solving for a desired output of a formula by changing its drivers. While these do not involve model building, they provide a level of functionality enabling what-if analysis, albeit with limited interactivity and expressivity.

\paragraph{BI Systems.} Commercial business intelligence (BI) tools recognize the need to augment business user's decision-making in data analysis systems. Both established players such as Power BI~\cite{ferrari2016introducing}, Tableau~\cite{tableaubusinessscience}, Salesforce~\cite{einsteindiscovery}, Facebook Robyn~\cite{robyn}, Alteryx~\cite{idoine2018magic}, SAS Visual Analytics~\cite{sas} and newcomers such as c3AI~\cite{c3aiexmachina}, Einblick~\cite{einblick}, and Sisu Data~\cite{sisudata} aim to provide various predictive and prescriptive functionalities. However, these tools either provide partial solutions whose effectiveness and adoption remain unclear or fail to cater to business users. \system specifically targets business users based on an intuitive set of interactive functions, providing a coherent workflow for data-driven business decision-making. 
\section{Conclusion}
\label{sec:conclusion}

In this paper, we start by defining the fundamental goal of business data analysis: improving business decisions based on data. This goal requires understanding and probing into the relationships between data variables representing hypothesized drivers and those representing key performance indicators (KPIs) that business users are interested in. We then introduce four functionalities for interactively learning and reasoning about these relationships. The success of model-driven analysis for broader audiences with no background in coding or data science depends on understanding the users and their goals and designing the user experience accordingly. It is not a technology-first problem. We, therefore, focus on providing intuitive interactive functionalities and provide a preliminary use case driven evaluation here. 

A close reading of Tukey's writings suggests that his emphasis on EDA and graphical representations was a reaction to (or an antithesis of) dry, purely confirmatory approaches to data analysis of his time. It is however time to bring the pendulum of data analysis currently skewed towards exploratory analysis to a synthesis reflecting the needs of large swaths of users in performing data analysis. This is important because the purpose of enterprise data analysis in practice on the ground is neither pretty pictures nor exploratory insights but improved decisions.

\bibliographystyle{ACM-Reference-Format}
\bibliography{main}


\begin{thebibliography}{40}


\ifx \showCODEN    \undefined \def \showCODEN     #1{\unskip}     \fi
\ifx \showDOI      \undefined \def \showDOI       #1{#1}\fi
\ifx \showISBNx    \undefined \def \showISBNx     #1{\unskip}     \fi
\ifx \showISBNxiii \undefined \def \showISBNxiii  #1{\unskip}     \fi
\ifx \showISSN     \undefined \def \showISSN      #1{\unskip}     \fi
\ifx \showLCCN     \undefined \def \showLCCN      #1{\unskip}     \fi
\ifx \shownote     \undefined \def \shownote      #1{#1}          \fi
\ifx \showarticletitle \undefined \def \showarticletitle #1{#1}   \fi
\ifx \showURL      \undefined \def \showURL       {\relax}        \fi
\providecommand\bibfield[2]{#2}
\providecommand\bibinfo[2]{#2}
\providecommand\natexlab[1]{#1}
\providecommand\showeprint[2][]{arXiv:#2}

\bibitem[\protect\citeauthoryear{AI}{AI}{2021}]%
        {c3aiexmachina}
\bibfield{author}{\bibinfo{person}{C3 AI}.} \bibinfo{year}{2021}\natexlab{}.
\newblock \showarticletitle{C3 AI Ex Machina}.
\newblock  (\bibinfo{year}{2021}).
\newblock
\urldef\tempurl%
\url{https://c3.ai/products/c3-ai-ex-machina/}
\showURL{%
\tempurl}


\bibitem[\protect\citeauthoryear{Amershi, Chickering, Drucker, Lee, Simard, and
  Suh}{Amershi et~al\mbox{.}}{2015}]%
        {amershi2015modeltracker}
\bibfield{author}{\bibinfo{person}{Saleema Amershi}, \bibinfo{person}{Max
  Chickering}, \bibinfo{person}{Steven~M Drucker}, \bibinfo{person}{Bongshin
  Lee}, \bibinfo{person}{Patrice Simard}, {and} \bibinfo{person}{Jina Suh}.}
  \bibinfo{year}{2015}\natexlab{}.
\newblock \showarticletitle{Modeltracker: Redesigning performance analysis
  tools for machine learning}. In \bibinfo{booktitle}{\emph{Proceedings of the
  33rd Annual ACM Conference on Human Factors in Computing Systems}}.
  \bibinfo{pages}{337--346}.
\newblock


\bibitem[\protect\citeauthoryear{Benesty, Chen, Huang, and Cohen}{Benesty
  et~al\mbox{.}}{2009}]%
        {benesty2009pearson}
\bibfield{author}{\bibinfo{person}{Jacob Benesty}, \bibinfo{person}{Jingdong
  Chen}, \bibinfo{person}{Yiteng Huang}, {and} \bibinfo{person}{Israel Cohen}.}
  \bibinfo{year}{2009}\natexlab{}.
\newblock \showarticletitle{Pearson correlation coefficient}.
\newblock In \bibinfo{booktitle}{\emph{Noise reduction in speech processing}}.
  \bibinfo{publisher}{Springer}, \bibinfo{pages}{1--4}.
\newblock


\bibitem[\protect\citeauthoryear{Breiman}{Breiman}{2001a}]%
        {breiman2001random}
\bibfield{author}{\bibinfo{person}{Leo Breiman}.}
  \bibinfo{year}{2001}\natexlab{a}.
\newblock \showarticletitle{Random forests}.
\newblock \bibinfo{journal}{\emph{Machine learning}} \bibinfo{volume}{45},
  \bibinfo{number}{1} (\bibinfo{year}{2001}), \bibinfo{pages}{5--32}.
\newblock


\bibitem[\protect\citeauthoryear{Breiman}{Breiman}{2001b}]%
        {breiman2001statistical}
\bibfield{author}{\bibinfo{person}{Leo Breiman}.}
  \bibinfo{year}{2001}\natexlab{b}.
\newblock \showarticletitle{Statistical modeling: The two cultures (with
  comments and a rejoinder by the author)}.
\newblock \bibinfo{journal}{\emph{Statistical science}} \bibinfo{volume}{16},
  \bibinfo{number}{3} (\bibinfo{year}{2001}), \bibinfo{pages}{199--231}.
\newblock


\bibitem[\protect\citeauthoryear{Brillinger}{Brillinger}{2002}]%
        {brillinger2002johnwtukey}
\bibfield{author}{\bibinfo{person}{David~R. Brillinger}.}
  \bibinfo{year}{2002}\natexlab{}.
\newblock \showarticletitle{John W. Tukey: His Life and Professional
  Contributions}.
\newblock \bibinfo{journal}{\emph{The Annals of Statistics}}
  \bibinfo{volume}{30}, \bibinfo{number}{6} (\bibinfo{year}{2002}),
  \bibinfo{pages}{1535--1575}.
\newblock
\urldef\tempurl%
\url{http://www.jstor.org/stable/1558729}
\showURL{%
\tempurl}


\bibitem[\protect\citeauthoryear{Brucato, Yadav, Abouzied, Haas, and
  Meliou}{Brucato et~al\mbox{.}}{2020}]%
        {brucato2020stochastic}
\bibfield{author}{\bibinfo{person}{Matteo Brucato}, \bibinfo{person}{Nishant
  Yadav}, \bibinfo{person}{Azza Abouzied}, \bibinfo{person}{Peter~J Haas},
  {and} \bibinfo{person}{Alexandra Meliou}.} \bibinfo{year}{2020}\natexlab{}.
\newblock \showarticletitle{Stochastic package queries in probabilistic
  databases}. In \bibinfo{booktitle}{\emph{Proceedings of the 2020 ACM SIGMOD
  International Conference on Management of Data}}. \bibinfo{pages}{269--283}.
\newblock


\bibitem[\protect\citeauthoryear{Cai, Vagena, Perez, Arumugam, Haas, and
  Jermaine}{Cai et~al\mbox{.}}{2013}]%
        {cai2013simulation}
\bibfield{author}{\bibinfo{person}{Zhuhua Cai}, \bibinfo{person}{Zografoula
  Vagena}, \bibinfo{person}{Luis Perez}, \bibinfo{person}{Subramanian
  Arumugam}, \bibinfo{person}{Peter~J Haas}, {and} \bibinfo{person}{Christopher
  Jermaine}.} \bibinfo{year}{2013}\natexlab{}.
\newblock \showarticletitle{Simulation of database-valued Markov chains using
  SimSQL}. In \bibinfo{booktitle}{\emph{Proceedings of the 2013 ACM SIGMOD
  International Conference on Management of Data}}. \bibinfo{pages}{637--648}.
\newblock


\bibitem[\protect\citeauthoryear{Cavallo and Demiralp}{Cavallo and
  Demiralp}{2018}]%
        {cavallo2018visual}
\bibfield{author}{\bibinfo{person}{Marco Cavallo} {and}
  \bibinfo{person}{{\c{C}}a{\u{g}}atay Demiralp}.}
  \bibinfo{year}{2018}\natexlab{}.
\newblock \showarticletitle{A visual interaction framework for dimensionality
  reduction based data exploration}. In \bibinfo{booktitle}{\emph{CHI}}.
  \bibinfo{pages}{1--13}.
\newblock


\bibitem[\protect\citeauthoryear{Cheng, Ming, and Qu}{Cheng
  et~al\mbox{.}}{2020}]%
        {cheng2020dece}
\bibfield{author}{\bibinfo{person}{Furui Cheng}, \bibinfo{person}{Yao Ming},
  {and} \bibinfo{person}{Huamin Qu}.} \bibinfo{year}{2020}\natexlab{}.
\newblock \showarticletitle{DECE: Decision Explorer with Counterfactual
  Explanations for Machine Learning Models}.
\newblock \bibinfo{journal}{\emph{IEEE Transactions on Visualization and
  Computer Graphics}} \bibinfo{volume}{27}, \bibinfo{number}{2}
  (\bibinfo{year}{2020}), \bibinfo{pages}{1438--1447}.
\newblock


\bibitem[\protect\citeauthoryear{Chuang, Ramage, Manning, and Heer}{Chuang
  et~al\mbox{.}}{2012}]%
        {chuang2012interpretation}
\bibfield{author}{\bibinfo{person}{Jason Chuang}, \bibinfo{person}{Daniel
  Ramage}, \bibinfo{person}{Christopher Manning}, {and}
  \bibinfo{person}{Jeffrey Heer}.} \bibinfo{year}{2012}\natexlab{}.
\newblock \showarticletitle{Interpretation and trust: Designing model-driven
  visualizations for text analysis}. In \bibinfo{booktitle}{\emph{Proceedings
  of the SIGCHI Conference on Human Factors in Computing Systems}}.
  \bibinfo{pages}{443--452}.
\newblock


\bibitem[\protect\citeauthoryear{Data}{Data}{2021}]%
        {sisudata}
\bibfield{author}{\bibinfo{person}{Sisu Data}.}
  \bibinfo{year}{2021}\natexlab{}.
\newblock \showarticletitle{Sisu Datas Blog}.
\newblock  (\bibinfo{year}{2021}).
\newblock
\urldef\tempurl%
\url{https://sisudata.com/}
\showURL{%
\tempurl}


\bibitem[\protect\citeauthoryear{Einblick}{Einblick}{2021}]%
        {einblick}
\bibfield{author}{\bibinfo{person}{Einblick}.} \bibinfo{year}{2021}\natexlab{}.
\newblock \showarticletitle{Einblick}.
\newblock  (\bibinfo{year}{2021}).
\newblock
\urldef\tempurl%
\url{https://einblick.ai/}
\showURL{%
\tempurl}


\bibitem[\protect\citeauthoryear{Ferrari and Russo}{Ferrari and Russo}{2016}]%
        {ferrari2016introducing}
\bibfield{author}{\bibinfo{person}{Alberto Ferrari} {and}
  \bibinfo{person}{Marco Russo}.} \bibinfo{year}{2016}\natexlab{}.
\newblock \bibinfo{booktitle}{\emph{Introducing Microsoft Power BI}}.
\newblock \bibinfo{publisher}{Microsoft Press}.
\newblock


\bibitem[\protect\citeauthoryear{Gale, Seiden, Atwood, Frantz, Woollen, and
  Çağatay Demiralp}{Gale et~al\mbox{.}}{2021}]%
        {gale2021sigma}
\bibfield{author}{\bibinfo{person}{James Gale}, \bibinfo{person}{Max Seiden},
  \bibinfo{person}{Gretchen Atwood}, \bibinfo{person}{Jason Frantz},
  \bibinfo{person}{Rob Woollen}, {and} \bibinfo{person}{Çağatay Demiralp}.}
  \bibinfo{year}{2021}\natexlab{}.
\newblock \bibinfo{title}{Sigma Worksheet: Interactive Construction of OLAP
  Queries}.
\newblock
\newblock
\showeprint[arxiv]{2012.00697}~[cs.DB]


\bibitem[\protect\citeauthoryear{Galhotra, Pradhan, and Salimi}{Galhotra
  et~al\mbox{.}}{2021}]%
        {galhotra2021explaining}
\bibfield{author}{\bibinfo{person}{Sainyam Galhotra}, \bibinfo{person}{Romila
  Pradhan}, {and} \bibinfo{person}{Babak Salimi}.}
  \bibinfo{year}{2021}\natexlab{}.
\newblock \showarticletitle{Explaining black-box algorithms using probabilistic
  contrastive counterfactuals}. In \bibinfo{booktitle}{\emph{Proceedings of the
  2021 International Conference on Management of Data}}.
  \bibinfo{pages}{577--590}.
\newblock


\bibitem[\protect\citeauthoryear{Gomez, Holter, Yuan, and Bertini}{Gomez
  et~al\mbox{.}}{2020}]%
        {gomez2020vice}
\bibfield{author}{\bibinfo{person}{Oscar Gomez}, \bibinfo{person}{Steffen
  Holter}, \bibinfo{person}{Jun Yuan}, {and} \bibinfo{person}{Enrico Bertini}.}
  \bibinfo{year}{2020}\natexlab{}.
\newblock \showarticletitle{ViCE: visual counterfactual explanations for
  machine learning models}. In \bibinfo{booktitle}{\emph{Proceedings of the
  25th International Conference on Intelligent User Interfaces}}.
  \bibinfo{pages}{531--535}.
\newblock


\bibitem[\protect\citeauthoryear{Haas, Maglio, Selinger, and Tan}{Haas
  et~al\mbox{.}}{2011}]%
        {haas2011data}
\bibfield{author}{\bibinfo{person}{Peter~J Haas}, \bibinfo{person}{Paul~P
  Maglio}, \bibinfo{person}{Patricia~G Selinger}, {and}
  \bibinfo{person}{Wang-Chiew Tan}.} \bibinfo{year}{2011}\natexlab{}.
\newblock \showarticletitle{Data is dead... without what-if models}.
\newblock \bibinfo{journal}{\emph{Proceedings of the VLDB Endowment}}
  \bibinfo{volume}{4}, \bibinfo{number}{12} (\bibinfo{year}{2011}),
  \bibinfo{pages}{1486--1489}.
\newblock


\bibitem[\protect\citeauthoryear{Hohman, Head, Caruana, DeLine, and
  Drucker}{Hohman et~al\mbox{.}}{2019}]%
        {hohman2019gamut}
\bibfield{author}{\bibinfo{person}{Fred Hohman}, \bibinfo{person}{Andrew Head},
  \bibinfo{person}{Rich Caruana}, \bibinfo{person}{Robert DeLine}, {and}
  \bibinfo{person}{Steven~M Drucker}.} \bibinfo{year}{2019}\natexlab{}.
\newblock \showarticletitle{Gamut: A design probe to understand how data
  scientists understand machine learning models}. In
  \bibinfo{booktitle}{\emph{Proceedings of the 2019 CHI conference on human
  factors in computing systems}}. \bibinfo{pages}{1--13}.
\newblock


\bibitem[\protect\citeauthoryear{Hullman and Gelman}{Hullman and
  Gelman}{2021}]%
        {hullman2021designing}
\bibfield{author}{\bibinfo{person}{Jessica Hullman} {and}
  \bibinfo{person}{Andrew Gelman}.} \bibinfo{year}{2021}\natexlab{}.
\newblock \showarticletitle{Designing for interactive exploratory data analysis
  requires theories of graphical inference}.
\newblock  (\bibinfo{year}{2021}).
\newblock


\bibitem[\protect\citeauthoryear{Idoine, Krensky, Brethenoux, Hare, Sicular,
  and Vashisth}{Idoine et~al\mbox{.}}{2018}]%
        {idoine2018magic}
\bibfield{author}{\bibinfo{person}{Carlie Idoine}, \bibinfo{person}{Peter
  Krensky}, \bibinfo{person}{Erick Brethenoux}, \bibinfo{person}{Jim Hare},
  \bibinfo{person}{Svetlana Sicular}, {and} \bibinfo{person}{Shubhangi
  Vashisth}.} \bibinfo{year}{2018}\natexlab{}.
\newblock \showarticletitle{Magic Quadrant for data science and
  machine-learning platforms}.
\newblock \bibinfo{journal}{\emph{Gartner, Inc}} (\bibinfo{year}{2018}),
  \bibinfo{pages}{13}.
\newblock


\bibitem[\protect\citeauthoryear{Keim, Andrienko, Fekete, G{\"o}rg, Kohlhammer,
  and Melan{\c{c}}on}{Keim et~al\mbox{.}}{2008}]%
        {keim2008visual}
\bibfield{author}{\bibinfo{person}{Daniel Keim}, \bibinfo{person}{Gennady
  Andrienko}, \bibinfo{person}{Jean-Daniel Fekete}, \bibinfo{person}{Carsten
  G{\"o}rg}, \bibinfo{person}{J{\"o}rn Kohlhammer}, {and} \bibinfo{person}{Guy
  Melan{\c{c}}on}.} \bibinfo{year}{2008}\natexlab{}.
\newblock \showarticletitle{Visual analytics: Definition, process, and
  challenges}.
\newblock In \bibinfo{booktitle}{\emph{Information visualization}}.
  \bibinfo{publisher}{Springer}, \bibinfo{pages}{154--175}.
\newblock


\bibitem[\protect\citeauthoryear{Louppe}{Louppe}{2017}]%
        {louppe2017bayesian}
\bibfield{author}{\bibinfo{person}{Gilles Louppe}.}
  \bibinfo{year}{2017}\natexlab{}.
\newblock \showarticletitle{Bayesian optimisation with Scikit-Optimize}.
\newblock  (\bibinfo{year}{2017}).
\newblock


\bibitem[\protect\citeauthoryear{Microsoft}{Microsoft}{2021a}]%
        {exceldocsolver}
\bibfield{author}{\bibinfo{person}{Microsoft}.}
  \bibinfo{year}{2021}\natexlab{a}.
\newblock \showarticletitle{Microsoft Excel Documentation}.
\newblock  (\bibinfo{year}{2021}).
\newblock
\urldef\tempurl%
\url{https://support.microsoft.com/en-us/office/define-and-solve-a-problem-by-using-solver-5d1a388f-079d-43ac-a7eb-f63e45925040/}
\showURL{%
\tempurl}


\bibitem[\protect\citeauthoryear{Microsoft}{Microsoft}{2021b}]%
        {exceldocgs}
\bibfield{author}{\bibinfo{person}{Microsoft}.}
  \bibinfo{year}{2021}\natexlab{b}.
\newblock \showarticletitle{Microsoft Excel Documentation}.
\newblock  (\bibinfo{year}{2021}).
\newblock
\urldef\tempurl%
\url{https://support.microsoft.com/en-us/office/use-goal-seek-to-find-the-result-you-want-by-adjusting-an-input-value-320cb99e-f4a4-417f-b1c3-4f369d6e66c7/}
\showURL{%
\tempurl}


\bibitem[\protect\citeauthoryear{Nickerson}{Nickerson}{1998}]%
        {nickerson1998confirmation}
\bibfield{author}{\bibinfo{person}{Raymond~S Nickerson}.}
  \bibinfo{year}{1998}\natexlab{}.
\newblock \showarticletitle{Confirmation bias: A ubiquitous phenomenon in many
  guises}.
\newblock \bibinfo{journal}{\emph{Review of general psychology}}
  (\bibinfo{year}{1998}).
\newblock


\bibitem[\protect\citeauthoryear{Pedregosa, Varoquaux, Gramfort, Michel,
  Thirion, Grisel, Blondel, Prettenhofer, Weiss, Dubourg,
  et~al\mbox{.}}{Pedregosa et~al\mbox{.}}{2011}]%
        {pedregosa2011scikit}
\bibfield{author}{\bibinfo{person}{Fabian Pedregosa}, \bibinfo{person}{Ga{\"e}l
  Varoquaux}, \bibinfo{person}{Alexandre Gramfort}, \bibinfo{person}{Vincent
  Michel}, \bibinfo{person}{Bertrand Thirion}, \bibinfo{person}{Olivier
  Grisel}, \bibinfo{person}{Mathieu Blondel}, \bibinfo{person}{Peter
  Prettenhofer}, \bibinfo{person}{Ron Weiss}, \bibinfo{person}{Vincent
  Dubourg}, {et~al\mbox{.}}} \bibinfo{year}{2011}\natexlab{}.
\newblock \showarticletitle{Scikit-learn: Machine learning in Python}.
\newblock \bibinfo{journal}{\emph{the Journal of machine Learning research}}
  \bibinfo{volume}{12} (\bibinfo{year}{2011}), \bibinfo{pages}{2825--2830}.
\newblock


\bibitem[\protect\citeauthoryear{Pirolli and Card}{Pirolli and Card}{2005}]%
        {pirolli2005sensemaking}
\bibfield{author}{\bibinfo{person}{Peter Pirolli} {and} \bibinfo{person}{Stuart
  Card}.} \bibinfo{year}{2005}\natexlab{}.
\newblock \showarticletitle{The sensemaking process and leverage points for
  analyst technology as identified through cognitive task analysis}. In
  \bibinfo{booktitle}{\emph{Proceedings of international conference on
  intelligence analysis}}.
\newblock


\bibitem[\protect\citeauthoryear{Robyn}{Robyn}{2021}]%
        {robyn}
\bibfield{author}{\bibinfo{person}{Robyn}.} \bibinfo{year}{2021}\natexlab{}.
\newblock \showarticletitle{facebookexperimental}.
\newblock  (\bibinfo{year}{2021}).
\newblock
\urldef\tempurl%
\url{https://facebookexperimental.github.io/Robyn/}
\showURL{%
\tempurl}


\bibitem[\protect\citeauthoryear{Roth}{Roth}{1988}]%
        {roth1988shapley}
\bibfield{author}{\bibinfo{person}{Alvin~E Roth}.}
  \bibinfo{year}{1988}\natexlab{}.
\newblock \bibinfo{booktitle}{\emph{The Shapley value: essays in honor of Lloyd
  S. Shapley}}.
\newblock \bibinfo{publisher}{Cambridge University Press}.
\newblock


\bibitem[\protect\citeauthoryear{Sanchez}{Sanchez}{2020}]%
        {Sanchez20}
\bibfield{author}{\bibinfo{person}{Susan~M. Sanchez}.}
  \bibinfo{year}{2020}\natexlab{}.
\newblock \showarticletitle{Data Farming: Methods for the Present,
  Opportunities for the Future}.
\newblock \bibinfo{journal}{\emph{{ACM} Trans. Model. Comput. Simul.}}
  \bibinfo{volume}{30}, \bibinfo{number}{4} (\bibinfo{year}{2020}),
  \bibinfo{pages}{22:1--22:30}.
\newblock


\bibitem[\protect\citeauthoryear{SAS}{SAS}{2021}]%
        {sas}
\bibfield{author}{\bibinfo{person}{SAS}.} \bibinfo{year}{2021}\natexlab{}.
\newblock \showarticletitle{SAS Visual Analytics}.
\newblock  (\bibinfo{year}{2021}).
\newblock
\urldef\tempurl%
\url{https://www.sas.com/en_us/software/visual-analytics.html}
\showURL{%
\tempurl}


\bibitem[\protect\citeauthoryear{Shang, Zgraggen, Buratti, Eichmann,
  Karimeddiny, Meyer, Runnels, and Kraska}{Shang et~al\mbox{.}}{2021}]%
        {shang2021davos}
\bibfield{author}{\bibinfo{person}{Zeyuan Shang}, \bibinfo{person}{Emanuel
  Zgraggen}, \bibinfo{person}{Benedetto Buratti}, \bibinfo{person}{Philipp
  Eichmann}, \bibinfo{person}{Navid Karimeddiny}, \bibinfo{person}{Charlie
  Meyer}, \bibinfo{person}{Wesley Runnels}, {and} \bibinfo{person}{Tim
  Kraska}.} \bibinfo{year}{2021}\natexlab{}.
\newblock \showarticletitle{Davos: a system for interactive data-driven
  decision making}.
\newblock \bibinfo{journal}{\emph{Proceedings of the VLDB Endowment}}
  \bibinfo{volume}{14}, \bibinfo{number}{12} (\bibinfo{year}{2021}),
  \bibinfo{pages}{2893--2905}.
\newblock


\bibitem[\protect\citeauthoryear{Tableau}{Tableau}{2021a}]%
        {einsteindiscovery}
\bibfield{author}{\bibinfo{person}{Tableau}.} \bibinfo{year}{2021}\natexlab{a}.
\newblock \showarticletitle{Einstein Discovery}.
\newblock  (\bibinfo{year}{2021}).
\newblock
\urldef\tempurl%
\url{https://www.tableau.com/products/add-ons/einstein-discovery}
\showURL{%
\tempurl}


\bibitem[\protect\citeauthoryear{Tableau}{Tableau}{2021b}]%
        {tableaubusinessscience}
\bibfield{author}{\bibinfo{person}{Tableau}.} \bibinfo{year}{2021}\natexlab{b}.
\newblock \showarticletitle{What is Tableau Business Science?}
\newblock  (\bibinfo{year}{2021}).
\newblock
\urldef\tempurl%
\url{https://www.tableau.com/about/blog/2021/3/what-is-tableau-business-science}
\showURL{%
\tempurl}


\bibitem[\protect\citeauthoryear{Tukey}{Tukey}{1962}]%
        {tukey1962future}
\bibfield{author}{\bibinfo{person}{John~W Tukey}.}
  \bibinfo{year}{1962}\natexlab{}.
\newblock \showarticletitle{The future of data analysis}.
\newblock \bibinfo{journal}{\emph{The annals of mathematical statistics}}
  \bibinfo{volume}{33}, \bibinfo{number}{1} (\bibinfo{year}{1962}),
  \bibinfo{pages}{1--67}.
\newblock


\bibitem[\protect\citeauthoryear{Tversky and Kahneman}{Tversky and
  Kahneman}{1974}]%
        {tversky1974judgment}
\bibfield{author}{\bibinfo{person}{Amos Tversky} {and} \bibinfo{person}{Daniel
  Kahneman}.} \bibinfo{year}{1974}\natexlab{}.
\newblock \showarticletitle{Judgment under uncertainty: Heuristics and biases}.
\newblock \bibinfo{journal}{\emph{Science}} (\bibinfo{year}{1974}),
  \bibinfo{pages}{1124--1131}.
\newblock


\bibitem[\protect\citeauthoryear{Wexler, Pushkarna, Bolukbasi, Wattenberg,
  Vi{\'e}gas, and Wilson}{Wexler et~al\mbox{.}}{2019}]%
        {wexler2019if}
\bibfield{author}{\bibinfo{person}{James Wexler}, \bibinfo{person}{Mahima
  Pushkarna}, \bibinfo{person}{Tolga Bolukbasi}, \bibinfo{person}{Martin
  Wattenberg}, \bibinfo{person}{Fernanda Vi{\'e}gas}, {and}
  \bibinfo{person}{Jimbo Wilson}.} \bibinfo{year}{2019}\natexlab{}.
\newblock \showarticletitle{The what-if tool: Interactive probing of machine
  learning models}.
\newblock \bibinfo{journal}{\emph{IEEE transactions on visualization and
  computer graphics}} \bibinfo{volume}{26}, \bibinfo{number}{1}
  (\bibinfo{year}{2019}), \bibinfo{pages}{56--65}.
\newblock


\bibitem[\protect\citeauthoryear{Zar}{Zar}{2005}]%
        {zar2005spearman}
\bibfield{author}{\bibinfo{person}{Jerrold~H Zar}.}
  \bibinfo{year}{2005}\natexlab{}.
\newblock \showarticletitle{Spearman rank correlation}.
\newblock \bibinfo{journal}{\emph{Encyclopedia of biostatistics}}
  \bibinfo{volume}{7} (\bibinfo{year}{2005}).
\newblock


\bibitem[\protect\citeauthoryear{Zhao, Wu, Lee, and Cui}{Zhao
  et~al\mbox{.}}{2018}]%
        {zhao2018iforest}
\bibfield{author}{\bibinfo{person}{Xun Zhao}, \bibinfo{person}{Yanhong Wu},
  \bibinfo{person}{Dik~Lun Lee}, {and} \bibinfo{person}{Weiwei Cui}.}
  \bibinfo{year}{2018}\natexlab{}.
\newblock \showarticletitle{iforest: Interpreting random forests via visual
  analytics}.
\newblock \bibinfo{journal}{\emph{IEEE transactions on visualization and
  computer graphics}} \bibinfo{volume}{25}, \bibinfo{number}{1}
  (\bibinfo{year}{2018}), \bibinfo{pages}{407--416}.
\newblock


\end{thebibliography}

\end{document}